\documentclass[journal=jacsat,manuscript=article]{achemso}

\usepackage[version=3]{mhchem} 
\usepackage{xcolor}

\author{Huatian Hu}
\affiliation[Wuhan Institute of Technology]{Hubei Key Laboratory of Optical Information and Pattern Recognition, Wuhan Institute of Technology, Wuhan 430205, China}
\alsoaffiliation[Istituto Italiano di Tecnologia]{Istituto Italiano di Tecnologia, Center for Biomolecular Nanotechnologies, Via Barsanti 14, 73010 Arnesano, Italy}

\author{Xin Shu}
\affiliation[Wuhan Institute of Technology]{Hubei Key Laboratory of Optical Information and Pattern Recognition, Wuhan Institute of Technology, Wuhan 430205, China}

\author{Zhiwei Hu}
\affiliation[East China Normal University]{State Key Laboratory of Precision Spectroscopy, East China Normal University, 200241 Shanghai, China}

\author{Di Zheng}
\affiliation[Shenzhen University]{\textit{State Key Laboratory of Radio Frequency Heterogeneous Integration} , Shenzhen University,518060 Shenzhen, China}

\author{Wei Dai}
\affiliation[Wuhan University]{School of Physics and Technology, Wuhan University, Wuhan 430072, China}

\author{Xiang Lan}
\affiliation[Istituto Italiano di Tecnologia]{Istituto Italiano di Tecnologia, Center for Biomolecular Nanotechnologies, Via Barsanti 14, 73010 Arnesano, Italy}

\author{Xiaobo Han}
\email{hanxiaobo@wit.edu.cn}
\affiliation[Wuhan Institute of Technology]{Hubei Key Laboratory of Optical Information and Pattern Recognition, Wuhan Institute of Technology, Wuhan 430205, China}

\author{Wen Chen}
\email{wchen@lps.ecnu.edu.cn}
\affiliation[East China Normal University]{State Key Laboratory of Precision Spectroscopy, East China Normal University, 200241 Shanghai, China}

\author{Hongxing Xu}
\email{hxxu@hnas.ac.cn}
\affiliation[Henan Academy of Sciences]
{Institute of Laser Manufacturing, Henan Academy of Sciences, Zhengzhou, China.}
\alsoaffiliation[East China Normal University]{State Key Laboratory of Precision Spectroscopy, East China Normal University, 200241 Shanghai, China}

\title[An \textsf{achemso} demo]
  {Extreme polaritonic interactions in a room-temperature deterministic sub-nanocavity quantum electrodynamic platform}

\abbreviations{IR,NMR,UV}

\begin{document}


\begin{abstract}

Pushing nanoscale optical confinement to its ultimate limits defines the regime of nano-cavity quantum electrodynamics (nano-cQED), where light--matter interactions approach the fundamental quantum limits of individual atoms, e.g., picocavities. However, realizing such extreme confinement in a stable and controllable manner remains a key challenge. Here, we introduce a van der Waals material–based nano-cQED platform by coupling monolayer MoS$_2$ excitons to plasmonic sub-nanocavities formed via assembly of ultrasmall gold clusters (3--5 nm) in the nanogap of a nanoparticle-on-mirror nanocavity. These clusters emulate the field-confining role of atomic protrusions of the picocavities through a resonance-insensitive lightning-rod effect, achieving deep-subwavelength mode volumes. In this nano-cQED testbed, we observe pronounced multi-branch Rabi splittings ($>200$ meV, $\frac{\Omega}{\sum\Gamma/2}\simeq2$) and ultrastrong lower-branch polaritonic photoluminescence with up to 10$^4$-fold enhancement. This deterministic architecture provides a controllable pathway to access picocavity-like behavior and opens new opportunities for single-molecule spectroscopy and the exploration of nano-cQED.

\end{abstract}

\section{Introduction}
Cavity quantum electrodynamics (cQED) studies fundamental interaction dynamics of matter with quantized light trapped in a cavity. 
The discovery that localized surface plasmon polaritons (LSPs)--quantum superposition of photons and free electrons--can further confine light into deep-subwavelength \textit{nanocavities} beyond the diffraction limit marked a turning point in extreme nanophotonics \cite{baumberg_extreme_2019}, which pushes conventional cQED into nano-cQED regime.
This extreme confinement squeezes the spatial distribution of the optical field \cite{xuElectromagneticContributionsSinglemolecule2000,savage_revealing_2012,ciraci_probing_2012,chen_probing_2018,zhang_switching_2022} to be comparable with a wide spectrum of single quantum quasi-particles. This enables nanoscale light–matter interactions in the quantum mechanical level from weak to strong regimes \cite{chikkaraddy_single-molecule_2016,hu2024robust,zheng2025active} with broad applications in single-molecule Raman \cite{nie_probing_1997,xuSpectroscopySingleHemoglobin1999a,jiang_distinguishing_2015}, fluorescence \cite{kinkhabwala_large_2009,yang_sub-nanometre_2020,lu_plexciton_2025}, and chirality sensing \cite{zhang_unraveling_2019,zhang_quantum_2024} , single-exciton polaritonics \cite{chikkaraddy_single-molecule_2016,liu_strong_2017,qin_revealing_2020,liu2024deterministic} and chemistry \cite{de_nijs_plasmonic_2017,xiang_molecular_2024}, hot electrons and thermal generation \cite{brongersma_plasmon-induced_2015,dubi_hot_2019}, plasmonic lasing \cite{azzam_ten_2020}, Bose-Eienstein condensation\cite{hakala_boseeinstein_2018}, and beyond. 

When a single or an ensemble of identical two-level systems interacts with a weakly excited quantized bosonic field, the Jaynes--Cummings or Tavis--Cummings model can be applied, respectively. \cite{larson2024jaynes} The coupling strength between \( N \) identical dipoles (each with transition moment \(\boldsymbol{\mu} = \mu_{12}\hat{\mathbf{d}}\)) and a vacuum field mode (\(\hat{\mathbf{f}}\)) of effective mode volume \(V_{\mathrm{m}}\) is characterized by the Rabi frequency \(\Omega = \mu_{12}\, \hat{\mathbf{d}} \cdot \sqrt{2N\omega_c/(\hbar\varepsilon_0 V_{\mathrm{m}})}\,\hat{\mathbf{f}}\). For a material with fixed intrinsic properties, tuning the optical field provides an effective means to modulate the light--matter interaction, which scales as \(\Omega \propto \sqrt{N/V_{\mathrm{m}}}\). Plasmonic systems, in particular, aim to minimize \(V_{\mathrm{m}}\) to achieve extreme coupling strengths and enhanced light--matter interactions~\cite{chikkaraddy_single-molecule_2016,urbieta_atomic-scale_2018,benzSinglemoleculeOptomechanicsPicocavities2016a}. Although increasing the number of two-level systems enhances the Rabi frequency as \(\Omega \propto \sqrt{N}\), \cite{kleemann2017strong,stuhrenberg2018strong} the presence of multiple dipoles tends to wash out quantum features and reduce the intrinsic nonlinearity, underscoring the importance of achieving strong coupling at the single-emitter level.

Since then, plasmonic nanostructures with ultrasmall dimensions—ranging from engineered nanomorphologies (gaps \cite{xuSpectroscopySingleHemoglobin1999a,ciraci_probing_2012,lei2012revealing}, corners \cite{barbry_atomistic_2015,liu_strong_2017}, and protrusions \cite{benzSinglemoleculeOptomechanicsPicocavities2016a,wu_bright_2021,li_bright_2021}) to two-dimensional electron gases \cite{hu_low-power_2024,berkmann_ultrastrong_2024} and low-dimensional and transdimensional materials \cite{koppens_graphene_2011,boltasseva_transdimensional_2019}—have continued to push the limits of optical mode volume down to sub-nanometre scales. 
Single nanoparticles with sharp corners or protrusions can confine light to volumes on the order of tens of nanometers cubed, either through the lightning-rod effect \cite{barbry_atomistic_2015} or by supporting resonant quasi-normal modes \cite{liu_strong_2017}. A more accessible and scalable approach, however, is to leverage a nanogap-antenna configuration to host such localized and finely tuned morphologies.
\cite{barbry_atomistic_2015,urbieta_atomic-scale_2018,wu_bright_2021,li_bright_2021,benzSinglemoleculeOptomechanicsPicocavities2016a}. 
Thanks to the host antenna that efficiently harvests light, such a small feature—typically regarded as a dark resonance—can be excited as a bright mode \cite{li_bright_2021,wu_bright_2021}. This enables further compression of the antenna-confined field into a volume smaller than \(1 \,\mathrm{nm}^3\), \cite{li_bright_2021,wu_bright_2021} forming what is known as a \textit{picocavity} that can sense the local vibronic signal from a single molecule \cite{benzSinglemoleculeOptomechanicsPicocavities2016a}. 
However, the practical utility of such picocavities remains elusive, as they arise from transient atomic-scale features—such as adatoms or atomic protrusions—that, although following certain formation rules~\cite{kerner2025optical}, can be spatiotemporally random and readily created or annihilated.
Their lifecycle is highly sensitive to environmental factors, including temperature, local chemistry, and laser-induced photoreactions.  
Although these atomic-scale hotspots offer fascinating potential for controlling quantum-level interactions and enabling site-specific chemistry, their spatiotemporal uncertainty makes them extremely difficult to control, trace, and systematically investigate.


In this work, inspired by the mode confinement, i.e., tiny $V_{\rm m}$, observed in protrusions (adatoms) within picocavities, we propose a deterministic nano-cQED platform with designed sub-nanometric hotspots for extreme nanophotonics. Specifically, we experimentally introduce ultrasmall metallic nanoparticles or clusters (3–5 \,nm) into the nanogap of a nanoparticle-on-mirror (NPoM) cavity to emulate the picocavity effect. Recently, a similar structure \cite{aghdaee2025optical} with much larger inner particles has highlighted the sensitivity and optical accessibility of the architecture. Our work thus arrives at a timely juncture, extending this emerging nanocavity platform to the room-temperature quantum electrodynamic regime. Our embedded clusters exhibit a small radius of curvature, enabling a resonance-insensitive lightning-rod effect that confines light into a deeply subwavelength mode volume of approximately 55 \,nm$^3$ ($\simeq1.6\times10^{-7}\lambda_0^3$). 
To verify this extreme confinement and establish the applicability of the cQED framework, a monolayer of MoS\(_2\)—a valley-excitonic transition-metal dichalcogenide (TMD)—was inserted into the nanogap to demonstrate polaritonic behavior.
Remarkably large Rabi frequencies (208 meV) were observed from the sub-nanocavity, approximately twice that of the conventional NPoM (its host nanocavity). The corresponding figure of merit, $\Omega/\Gamma = 1.02$, considerably exceeds the typical range of 0.4–0.7 reported in previous studies by almost two times.
Extremely enhanced (25,000-fold) lower-branch polaritonics photoluminescence (PL) was recorded, confirming a strong Purcell enhancement. 
Statistical measurements over hundreds of structures revealed that the presence of the sub-nanocavity can yield up to tenfold stronger PL compared to standard NPoM cavities. In addition, based on quantum calculations, we propose that the sub-nanocavity can exhibit quantum light antibunching emission with second-order correlation function $g^{(2)}\simeq0.16$, suggesting its potential as a nano-cQED testbed. 

Unlike adatom-based picocavities, which form stochastically and are challenging to reproduce, our nanoparticle-cluster-on-mirror (NPcoM) platform offers deterministic, stable, and reproducible sub-nanometer hotspots. This approach opens new avenues for controlling light--matter interactions at the quantum level, with broad implications for single-exciton strong coupling, single-molecule spectroscopy, and quantum nano-optics.

\section{Results and discussion}
\begin{figure}[!hb]
    \centering
    \includegraphics[keepaspectratio, width=1\textwidth]{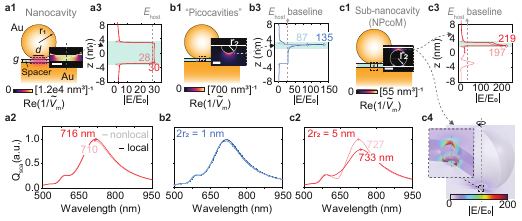}
    \caption{\footnotesize Schematics (\textbf{a1, b1, c1}), scattering spectra (\textbf{a2, b2, c2}) and field enhancements (\textbf{a3, b3, c3, c4}) of three different configurations: (\textbf{a1-3}) 100 nm nanoparticles-on-mirror (NPoM) \textit{nanocavity}, (\textbf{b1-3}) 100 nm NPoM with 1 nm diameter semispherical protrusion, i.e., emulating the \textit{picocavity}. (\textbf{c1-4}) 100 nm NPoM hosting a small gold cluster inside the gap with 5 nm diameter, forming a {NPcoM} \textit{sub-nanocavity}. In panels \textbf{a3}, \textbf{b3}, and \textbf{c3}, the blue regions schematically highlight the dimension of the electromagnetic hotspots, indicating a pronounced field confinement within the picocavity and sub-nanocavity cases. The near-uniform field enhancement ($E_{\rm host}$) in a nanocavity (\textbf{a3}) explains the electric field baselines in (\textbf{b3} and \textbf{c3}).
    Local-response model (solid lines) and nonlocal model (hydrodynamic theory, lighter lines) are compared for all results. 
    The gold protusions and seeds (i.e., clusters) are assumed on the axis to implement cylindrical coordinates. \textbf{c4} shows the 3D field enhancement of the NPcoM. Scalebars represent 10 nm (\textbf{a1}), 0.5 nm (\textbf{b1}), and 2.5 nm (\textbf{c1}), respectively.}   
    \label{fig:1}
\end{figure}

Nanocavities formed by a nanoparticle drop-cast or self-assembled on a metal substrate spaced by certain materials have been widely studied in the past decades due to their easily-fabricated and well-defined nanogaps and their controllable and outstanding optical performances. \cite{} 
Its exceptional modal brightness and light concentration can couple light with fine atomic features in the nanocavity. 
Based on these, NPoM configurations are ideal \textit{hosts} for creating picocavities \cite{}, with adatoms or protrusions like a \textit{guest}.
Thus, we start from investigating the host nanocavity as a starting point as well as a reference for comparison (Figs.~\ref{fig:1}a1-3).

As shown by the schematic in Fig.~\ref{fig:1}a, a 100 nm nanoparticle (radius $r_1=50$ nm) with $d=20$ nm flat bottom facet can concentrate light into the nanogaps under the facet. When the gap is as narrow as 5.8 nm (we set this value considering hosting a gold seed and a single layer of TMD, later on), a prominent $l_{01}$ mode \cite{tserkezis2015hybridization} is expected at 716 nm in the scattering spectra excited by an 80 deg oblique p-polarized light (Fig.~\ref{fig:1}b). A strong electric field enhancement of 30 times (or 28 times under the nonlocal model) is expected. See methods in Supporting Information (SI) for the details. This amount of field enhancement arises from the nanogap of the NPoM, which acts as a host with baseline enhancement $E_{\rm host}$ for any guest morphologies, such as a protrusions-supported picocavity, as shown in Figs.~\ref{fig:1}b1-3, and a small cluster in  Figs.~\ref{fig:1}c1-4.

The introduction of a small protrusion (0.5 or 1~nm) does not noticeably alter the far-field scattering spectrum, as shown in Figs.~\ref{fig:1}b2, due to the dimensional mismatch. In the near-field, however, considering a protrusion of size \(2r_2 = 1\)~nm, the electric field can be trapped by the small curvature, resulting in a 135-fold enhancement using the local model and an 87-fold enhancement using the nonlocal model. This large discrepancy is expected due to the extreme curvature~\cite{ciraci2013hydrodynamic} and its resultant quantum effect. 

In contrast, when a 5~nm Au seed is inserted into the nanogap (Fig.~\ref{fig:1}c1), the system exhibits both curvature-trapped behavior and the formation of a tiny nanogap (0.8~nm) with the top facet. This configuration provides additional enhancement, reaching 219-fold with the local model and 197-fold with the nonlocal model. Interestingly, due to this additional enhancement, the light shows an even tighter confinement in our NPcoM sub-nanocavity. The absence of a tiny curvature, \(2r_2 = 5\)~nm in the NPcoM system compared with the \(2r_2 = 1\)~nm picocavity, helps the field enhancement persist against quantum nonlocal effects. For clarity, we plot the full 3D field enhancement of the NPcoM sub-nanocavity in Fig.~\ref{fig:1}c4.
The NPcoM sub-nanocavity's far-field scattering (Fig. \ref{fig:1}c2) does present a red shift of the $l_{01}$ main peak from 716 nm to 733 nm compared with the host nanocavity, the nonlocal effect blueshifts it back to 727 nm. Therefore, the introduction of an extra 5 nm cluster does not noticeably shift the resonance. 
Finally, we examine the effective mode volume \(V_\mathrm{m}\) (see Methods in SI) of the NPcoM sub-nanocavity, which can be reduced to as small as 55~nm\(^3\).

\begin{figure}[!ht]
    \centering
    \includegraphics[keepaspectratio, width=1\textwidth]{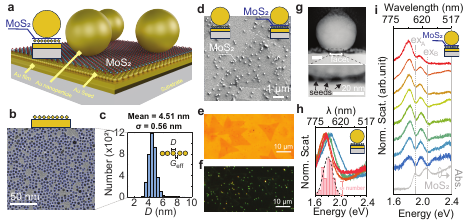}
    \caption{\footnotesize(\textbf{a})~3D schematic of the NPcoM structure, with the inset showing its 2D cross-section. 
(\textbf{b})~A representative TEM image of densely assembled 5~nm Au nanoseeds, with their size distribution shown in (\textbf{c}).  For the full TEM image from which the data in (\textbf{c}) are extracted, see SI Fig.~S3.
(\textbf{d})~SEM images of NPcoM structures on and off monolayer MoS$_2$, with corresponding schematics shown as insets. 
From bottom to top, the layers are Si substrate, 80 nm Au film, 5~nm nanoseeds, with or without a monolayer MoS$_2$, and a 100~nm Au nanoparticle. 
(\textbf{e}, \textbf{f})~Bright- and dark-field optical images of the same NPcoM region. 
(\textbf{g})~TEM images of the bare NPcoM sub-nanocavity's vertical cross-section confirming the presence of nanoclusters beneath the nanoparticle. The scalebars represent 20 nm. 
(\textbf{h}, \textbf{i})~Scattering spectra of NPcoM without (\textbf{h}) and with (\textbf{i}) a monolayer MoS$_2$ excitons, respectively. The absorption of the monolayer MoS$_2$ in (\textbf{i}) shows two peaks of A and B excitons.
}   
\label{fig:2}
\end{figure}

To deterministically realize the sub-nanocavity proposed in Fig.~\ref{fig:1}c1, a single layer of closely packed 5~nm Au nanoseeds (purchased from nanoComposix) was first prepared on the Au film by the interfacial self-assembly method, see Methods in SI~\cite{selfassem,supp}. 
Subsequently, CVD-grown monolayer MoS$_2$ (purchased from SixCarbon Tech., Shenzhen) was transferred onto the nanoclusters, followed by the deposition of 100~nm Au nanoparticles on top (schematic shown in Fig.~\ref{fig:2}a, see Methods). 
Transmission electron microscopy (TEM) characterization of the close-packed nanoseed assemblies on carbon films in Fig.~\ref{fig:2}b revealed a well-ordered monolayer of nanoseeds with high packing density, demonstrating uniform distribution free from aggregation or macroscopic defects. 
We used an in-house recognition algorithm \cite{zheng2023toward,zheng2024tunable} to perform the statistical analysis on the closely packed nanoseeds (see Methods). 
As shown in Fig.~\ref{fig:2}c, the nanoseed sizes follow a normal distribution with a mean of 4.51~nm and a standard deviation of 0.56~nm. The average lateral nanogap between adjacent nanoseeds is approximately 2.3~nm. 
From the TEM image, we also estimate a monolayer assembly yield with an average coverage ratio of 34\%. (see Methods in SI)

The scanning electron microscopy (SEM) image in Fig.~\ref{fig:2}d clearly reveals the geometries of the NPcoM structures located on and off the monolayer MoS$_2$ outlined by the triangular region with dashed lines. Bright- and dark-field images of the same region in Figs.~\ref{fig:2}e and \ref{fig:2}f demonstrate the yield of sub-nanocavity formation. The color of the Airy disks in Fig.~\ref{fig:2}f further confirms the uniformity of the cavities. The TEM image in Fig.~\ref{fig:2}g characterizes the vertical cross section of the NPcoM, from which it is evident that sub-nanocavities are formed beneath each 100~nm nanoparticle with an estimated facet size of 20~nm, though the number of sub-nanocavities can be randomly dependent on the localized assembly of the nanoseeds. This indeed gives inherent randomness to the light-matter interaction.

Fig.~\ref{fig:2}h presents representative single-particle dark-field scattering spectra of NPcoM structures off the MoS$_2$, corresponding to bare sub-nanocavities without embedded excitons (see Methods). The typical spectra are shown, with the distribution of resonance wavelengths summarized in a histogram derived from 52 nanoparticles. The mean resonance wavelength is centered at 690 nm, with a standard deviation of 30 nm. This considerable deviation arises from the uncertainty of the assembled nanoseeds underneath, as shown in Fig. \ref{fig:2}b, and the size fluctuation of 100 nm nanoparticles. 

Importantly, as the NPcoM sub-nanocavity couples with the monolayer MoS$_2$ excitons inside to form a nano-cQED platform at room temperature, a multibranch Rabi splitting is demonstrated as shown in Fig. \ref{fig:2}i. The gray curve shows the measured absorption of the monolayer MoS$_2$ arising from the A and B excitons (ex$\rm _A$ and ex$\rm _B$, respectively), which originate from spin--orbit-split valence bands at the K and K$'$ valleys of the Brillouin zone, corresponding to transitions from the upper and lower valence bands to the conduction band, respectively. As shown in Fig.~\ref{fig:2}i, the A exciton of monolayer MoS$_2$ is located at 1.89 eV, whereas the B exciton appears at 2.05 eV. Two pronounced energy splittings appear at the A- and B-exciton resonances, revealing three newly formed polaritonic eigenmodes that constitute a typical three-mode nano-cQED system. Its corresponding Hamiltonian can be written as

\begin{equation}
\hat{H} =
\begin{pmatrix}
\hbar\tilde{\omega}_{\mathrm{c}} & g_{\mathrm{1}} & g_{\mathrm{2}} \\
g_{\mathrm{1}} & \hbar\tilde{\omega}_{\mathrm{A}} & 0 \\
g_{\mathrm{2}} & 0 & \hbar\tilde{\omega}_{\mathrm{B}}
\end{pmatrix},
\label{eq:3mode_H}
\end{equation}

\noindent where $\tilde{\omega}_{j} = \omega_{j} - i\Gamma_{j}$ $[j = \mathrm{c\,(cavity), A, B}]$ denote the complex mode frequencies of plasmons, A and B excitons, respectively, whose imaginary parts represent the damping rates. $g_{\mathrm{1}}$ and $g_{\mathrm{2}}$ represent their respective coupling strengths. The diagonalization of the Hamiltonian matrix gives rise to the eigen-frequencies of the hybridized branches.

While scattering spectra are widely used to identify and quantify the coupling strength in plasmonic cQED platforms, their pronounced dips can also arise from weak or intermediate interactions, such as Fano interference or induced transparency \cite{hu2021unified}. These features originate from the coherent interference between the scattered and incident light. In contrast, photoluminescence (PL) is an incoherent process, in which excited electrons undergo relaxation that randomizes their phases, leading to emission at lower energies than the excitation. This makes PL a reliable indicator of Rabi splitting and strong coupling; namely, if the polaritonic PL emission appears at the hybridized branches of a cQED system, strong coupling can be unambiguously confirmed \cite{hu2021unified,niu2022unified}. 

\begin{figure}[!ht]
    \centering
    \includegraphics[keepaspectratio, width=1\textwidth]{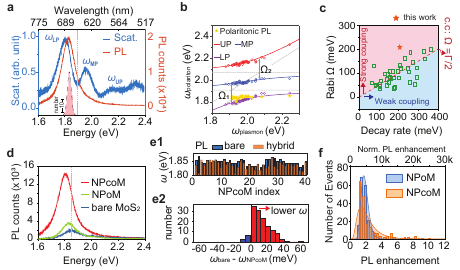}
    \caption{\footnotesize\textbf{(a)} Example of scattering and PL from the NPcoM exciton–polariton, showing three distinct branches. The statistical distribution of the enhanced PL peak positions is overlaid. \textbf{(b)} Multibranch anti-crossing dispersion with two Rabi splittings, $\Omega_1 = 208$ meV and $\Omega_2 = 210$ meV, fitted from the polaritonic frequency gathered from different nanoparticles. \textbf{(c)} Comparison of our results with the literature of plasmon-monolayer TMDC hybrid systems. The referred literature is summarized in Table S1 in the SI. \textbf{(d)} Enhanced polaritonic PL from a single NPcoM sub-nanocavity (red), and from an NPoM nanocavity (green). \textbf{(e1)} The histogram of peak locations of the PL from NPcoM-MoS$_2$ hybrid (orange) and the respective bare MoS$_2$ (blue) nearby. See full statistics in SI Fig. S7. The PL from the hybrid is generally redshifted relative to the bare MoS$_2$ nearby (see \textbf{(e2)} for the statistical differences in frequency), manifesting as a mix of lower-branch polaritonic PL. \textbf{(f)} Statistical analysis of PL enhancement across multiple devices illustrates the variability of NPcoM systems and their potential for realizing ultrabright light sources.}
\label{fig:3}
\end{figure}

In Fig.~\ref{fig:3}a, we demonstrate a representative example of polaritonic PL, hybridized from the plasmon and bright A excitons, which clearly deviates from the bare A-exciton energy (also see Figs. \ref{fig:3}d-e for detailed discussions) and aligns with the lower branch of the scattering spectrum. See Methods for PL measurements.
The emission is attributed to polaritonic PL, as it includes significant contributions from plasmon--exciton hybrid polaritons alongside residual uncoupled exciton emission from areas outside the nanoparticles.
Only the lower polaritonic branch is observed in PL owing to the rapid relaxation of the upper branch, a phenomenon widely reported in plasmon--exciton systems~\cite{melnikau2016rabi,liu2021nonlinear}. 
The statistical distribution of the PL emission peak positions is shown in Fig.~\ref{fig:3}a, exhibiting a normal distribution with a mean wavelength of 1.834 eV (675 nm) and a standard deviation of 0.014 eV.

We analyzed the polaritonic multibranches using an anti-crossing model, as shown in Fig.~\ref{fig:3}b. In addition to the scattering spectra, the peak location of the polaritonic PL was overlaid for direct comparison. The eigenvalues of the hybrid modes were obtained by diagonalizing the Hamiltonian of the cQED system defined by Eq.~(\ref{eq:3mode_H}) (see Methods for details for fitting), and the experimental data can be fitted to obtain the three complex eigenfrequencies $\omega_\mathrm{UP,MP,LP}$ representing the upper, middle, and lower plasmon-exciton polaritonic branches~\cite{nanophotonics2022}. From the fittings, the Rabi frequencies $\Omega_1=2g_1=208 $ meV, and $\Omega_2=2g_2=210 $ meV. According to the rigorous criteria for strong coupling, the observed Rabi splitting, $\Omega_1$, between A excitons and plasmons clearly satisfies the condition  $\Omega_1>(\Gamma_c+\Gamma_A)/2=102 $ meV, providing definitive evidence of strong coupling. Notably, the Rabi frequency exceeds twice the sum of the full widths at half maximum (FWHM) of two bare modes.
For the B exciton-plasmon system, the observed Rabi splitting, $\Omega_2$, similarly satisfies the strong coupling criterion  $\Omega_2>(\Gamma_c+\Gamma_B)/2=120 $ meV, confirming the establishment of strong coupling in this coupled system as well.

We also fitted individual scattering spectra using the coupled-oscillator model (see Methods and examples in SI). As summarized in Fig. S5, the mean Rabi splittings are $\Omega_1 = 2g_1 = 223 ~\mathrm{meV}$ and $\Omega_2 = 2g_2 = 222 ~ \mathrm{meV}$, with standard deviations of 37 meV and 34 meV, respectively, which meets aligns with the fitting from the dispersion in Fig. \ref{fig:3}b. The variations primarily originate from nanoscale morphological differences among the sub-nanocavity NPcoM devices. The maximum coupling strength reaches up to 300~meV, corresponding to approximately 15\% of the eigenmode frequency. 

As a reference, we fabricated its \textit{host} nanocavity without nanoseeds as a contrast sample—namely, 100-nm NPoM structure with an Al$_2$O$_3$ spacer and an embedded monolayer MoS$_2$ (see SI). As shown in Fig.~S6, only a single anticrossing between the plasmon and the A excitons was observed, without any evidence of multibranch coupling. The corresponding Rabi splitting of 130 meV is nearly half of that in our sub-nanocavity system. This reduction can be attributed to the significantly larger mode volume in the NPoM cavity (cf. Figs.~\ref{fig:1}a1 and \ref{fig:1}c1). Benefiting from the highly localized property of plasmonic nanocavities, strong coupling with TMDC materials can be achieved even at room temperature. SI Table S1 summarizes the key parameters of the monolayer TMDC exciton-plasmon hybrid systems, which are further visualized in Fig. \ref{fig:3}c. The results in Fig. \ref{fig:3}c show that the Rabi splitting energy of our system approaches the highest values reported to date. To account for the influence of losses on the splitting, we considered the ratio of coupling strength to decay, critical\;criteria $\rm{c.c} = 2\Omega/\Gamma=1$, with the red dashed line indicating the threshold for strong coupling ($\Omega/\Gamma=1/2$). In previously reported plexcitonic systems, this ratio typically ranges from 0.4 to 0.7. Remarkably, in our system, the ratio exceeds 1, demonstrating that our platform achieves an exceptionally high level of strong-coupling performance.

The NPoM can also exhibit relatively strong coupling, since the collective coupling strength scales as $ g \propto \sqrt{N/V_\mathrm{m}} $. The larger mode volume $V_\mathrm{m}$ in the NPoM allows many more excitons to participate, whereas the single-exciton coupling strength $ g_0 \propto 1/\sqrt{V_\mathrm{m}} $ is correspondingly smaller. However, for a cQED system, single- or few-exciton strong coupling is of particular interest because of the unique quantum effects it can sustain.  
Considering an average nanoseed diameter of 4.5~nm and a mean gap spacing of 2~nm (6.5~nm period, see Figs. \ref{fig:2}b, c), a 100-nm-diameter NPoM cavity with a 20-nm facet can accommodate approximately 9.5 nanoseeds (i.e., sub-nanocavities). The combined effective mode volume of these 9.5 sub-nanocavities accounts for only about 4.3\% of the $V_\mathrm{m}$ of the host NPoM nanocavity. Consequently, the NPcoM cavity contains roughly $1/23$ of the total exciton number $N$ compared with its host NPoM structure.  

Extreme field confinement (i.e, \textit{picocavities}) from atomic protrusions is known for randomly generating strong optical signals at a certain frequency of events \cite{benzSinglemoleculeOptomechanicsPicocavities2016a,kerner2025optical,li_bright_2021,wu_bright_2021}. Previous work \cite{chen_intrinsic_2021} on intrinsic metallic PL blinking from plasmonic junctions was also attributed to the formation of the Au clusters at the metallic surface with illumination. Interestingly, we observed similar extremely enhanced polaritonic PL from NPcoM as compared in Figs. \ref{fig:3}d. Fig. \ref{fig:3}d presents two cases of PL and polaritonic PL without and with NPoM nanocavity and NPcoM sub-nanocavity, respectively. Their statistical analysis is shown in Figs. \ref{fig:3}e-f. The PL enhancement factor, $I/I_0$, is defined relative to the PL intensity $I_0$ of bare MoS$_2$ on the substrate adjacent to the nanoparticle used for the polaritonic PL measurement. 
A roughly 2-fold enhancement is observed from the NPoM nanocavity, whereas the sub-nanocavity can support a PL up to 10-fold enhancement. 
Since the cavities in both cases have deep-subwavelength footprints compared with the incident beam spot, the PL enhancement should be normalized by the area factor $A_{\rm spot}/A_0=(1000/20)^2=2500$, where the diameter of the laser beam spot is $D_{\rm spot} \approx 1~\mu\mathrm{m}$, and 20 nm is the facet of the 100 nm nanoparticle according to Fig. \ref{fig:2}g.  Therefore, accounting for the area normalization, a factor of 2500 is expected for the cavities, which corresponds to a maximum PL enhancement of approximately 25,000-fold. 
 
We measured the photoluminescence (PL) from more than 140 NPcoMs and reference NPoMs, as summarized in Figs.~\ref{fig:3}e--f and SI Fig.~S7. To ascertain that the observed PL enhancement arises from strong coupling (polaritonic PL) rather than weak coupling in the Purcell regime, we analyzed the spectral positions of the PL with and without the nanostructures. Figure~\ref{fig:3}e1 presents the PL energies from first 40 NPcoMs together with the corresponding bare TMDC PL measured in their vicinity, with full statistics provided in SI Fig.~S7.
Because uncoupled excitons within the laser spot (outside the cavity) unavoidably contribute to the detected signal unless removed or quenched \cite{zheng2025active,lo2022plasmonic}, the measured PL consists of a superposition of the bare MoS$_2$ excitonic PL and the lower polaritonic branch. This superposition leads to an experimental redshift relative to the intrinsic A-exciton energy. As shown in Fig.~\ref{fig:3}e1 and SI Fig.~S7, the PL from all NPcoMs is consistently redshifted with respect to bare MoS$_2$ (see Figure~\ref{fig:3}e2 for the binned energy differences between PL with and without NPcoM sub-nanocavities), ruling out shifts induced by detuned weak plasmon-enhanced PL. We therefore attribute the PL measured from NPcoM--MoS$_2$ devices to emission from the lower polaritonic branch.

Similar to atomic-scale picocavities, the small nanoclusters in the NPcoM sub-nanocavities exhibit substantial variability in performance. This is due to the uncertain arrangement of the nanoseeds underneath, they were in sharp contrast to the relatively homogeneous but weaker emission observed from the NPoM nanocavity counterparts (Fig. \ref{fig:3}f). 
The NPoMs, despite slight fluctuations, show a near-normal distribution with a mean PL non-normalized enhancement of about twofold. A few low-probability events exhibit enhancements up to four-fold, likely originating from local morphological irregularities such as sharp edges or corners, as discussed in Ref.~\cite{leng2018strong}. In contrast, the NPcoMs display a much broader distribution extending beyond ten-fold enhancement, with a significant number of events exceeding the maximum observed in NPoM nanocavities. 

To benchmark the performance of our NPcoM platform against previously reported systems (also comprehensively reviewed in Ref.~\cite{wei_plasmonexciton_2021}), we summarize the comparison in SI Table~S2.
While most previous studies operated in the weak coupling (Purcell) regime, our NPcoM system functions under strong coupling conditions, achieving up to a 25,000-fold enhancement—comparable to the maximum enhancement reported for excitation–emission dual-resonant systems with well-aligned out-of-plane excitons in the weak-coupling regime \cite{B8}. Notably, our work represents the highest PL enhancement working in the strong coupling regime. 

\begin{figure}[!ht]
    \centering
    \includegraphics[keepaspectratio, width=1\textwidth]{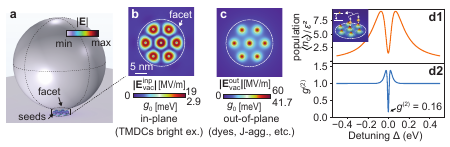}
    \caption{\footnotesize
    \textbf{(a)} Three-dimensional geometry of the NPcoM structure consisting of seven compact nanoseeds assembled inside a nanocavity formed by the facet. 
    Vacuum electric field distributions are shown for the \textbf{(b)} in-plane and \textbf{(c)} out-of-plane components. 
    The in-plane component can couple with in-plane excitons, such as the bright exciton ex\textsubscript{A} in MoS\textsubscript{2} ($\boldsymbol{\mu} = 7.36$ D \cite{yang2022strong}), 
    while the out-of-plane component couples efficiently with emitters possessing a vertical dipole moment, such as molecules or J-aggregates ($\boldsymbol{\mu} = 33.6$ D \cite{liu_strong_2017}). Single-exciton coupling strength $g_0=\mathbf{E}_{\rm vac}\cdot\boldsymbol{\mu}$ with MoS\textsubscript{2} \textbf{(b)} and J-aggregates \textbf{(c)} are shown, respectively.
    Photon blockade behavior is demonstrated in \textbf{(d)}, where two J-aggregate excitons coupled to the NPcoM sub-nanocavity photons exhibit 
    (\textbf{d1}) mode splitting in the normalized population and (\textbf{d2}) anti-bunching behavior in the second-order correlation function $g^{(2)}$.}
    \label{fig:4}
\end{figure}

To illustrate its relevance in cQED applications, we theoretically investigate the photon blockade effect and its role in single-photon emission. Based on the statistical analysis in Fig.~\ref{fig:2}b, the Au nanoseeds are taken to have a diameter of 4.5~nm, with effective interparticle gaps of 2.3~nm. Under a 20~nm facet, the NPoM nanocavity can therefore accommodate seven nanoseeds, as shown in Fig.~\ref{fig:4}a. For computational efficiency, nanoseeds outside the nanocavity are neglected, as the optical energy is predominantly confined within the nanocavity region \cite{chikkaraddy_single-molecule_2016}.

In the cQED framework, plasmonic photons couple to excitons through vacuum fluctuations, with the single-exciton coupling strength defined as $g_0 = (\boldsymbol{\mu} \cdot \mathbf{E}_{\mathrm{vac}})/\hbar$. The vacuum field $\mathbf{E}_{\mathrm{vac}}$ is obtained via quasinormal-mode (QNM) analysis~\cite{wu2023modal}, where the mode field is normalized such that the total electromagnetic energy equals $\hbar \omega_{\mathrm{c}}/2$. The QNM approach, extensively used for open optical cavities~\cite{wu2023modal,wu2021nanoscale,wu_bright_2021}, is implemented here using a customized model based on the MAN package~\cite{wu2023modal}. 
Figure~\ref{fig:4}b shows the in-plane component of this QNM's $\mathbf{E}_{\mathrm{vac}}$ at the center plane of the TMDC as experimentally illustrated in Figs. \ref{fig:2} and \ref{fig:3}, displaying seven donut-shaped hotspots centered at the nanoseeds. The field minima correspond to the nanoseed axes, where the field is predominantly vertical, leading to a weak in-plane component. The maximal in-plane vacuum field reaches 19~MV/m. Coupling this field with MoS\textsubscript{2} A excitons  ($\boldsymbol{\mu} = 7.36$ D \cite{yang2022strong}) yields a single-exciton coupling strength of $g_0 = 2.9$~meV, approximately five times larger than that achieved in conventional nanocavity–monolayer TMDC systems~\cite{zheng2017manipulating,sun2018light}.

In contrast, the out-of-plane component of the vacuum field is approximately three times larger than the in-plane component, reaching 60~MV/m at the center of the nanoseed axes. Emitters with a vertical dipole moment, such as J-aggregates~\cite{liu_strong_2017} and dye molecules~\cite{chikkaraddy_single-molecule_2016}, can efficiently couple to this field. Here, as an example, we consider an NPcoM coupled to a J-aggregate with a strong dipole moment of 33.6~D, experimentally validated for single-exciton--level strong coupling \cite{liu_strong_2017}. With this optimized design, the resulting single-exciton coupling strength reaches $g_0 = 41.7$~meV (Fig.~\ref{fig:4}c), which can be strong enough for investigating cQED effects.

For example, we consider that there are two excitons residing inside the NPcoM with coupling strength $g_0 = 41.7$~meV. The two-exciton system in Fig.~\ref{fig:4}d1 inset is chosen for simpler experimental implementation in the future. As schematically shown in the inset of Fig.~\ref{fig:4}d1, the two excitons (two-level systems) interact with the same plasmon mode in the NPcoM through a plasmon-exciton interaction with strength $g_i$ or $g_j$. The total Hamiltonian is given by $
H = H_0 + H_{\mathrm{int}} + H_{\mathrm{pump}}$,
where $H_0$ is the uncoupled Hamiltonian, $H_{\mathrm{int}}$ describes the interactions, and $H_{\mathrm{pump}}$ accounts for driving by an external field.
Neglecting decoherence, the uncoupled Hamiltonian reads
\[
H_0 = \hbar \omega_0 \left(\sigma_i^\dagger \sigma_i + \sigma_j^\dagger \sigma_j\right) + \hbar \omega_{\mathrm{c}} b^\dagger b,
\]
where $\hbar \omega_0$ and $\hbar \omega_{\mathrm{c}}$ are the energies of the excitons and the cavity plasmons, respectively. $\sigma_{i,j}\ (\sigma_{i,j}^\dagger)$ are fermionic annihilation (creation) operators for the $i$-th and $j$-th excitons, while $b\ (b^\dagger)$ is the bosonic annihilation (creation) operator of the single-mode plasmon.

The interaction Hamiltonian includes both plasmon-exciton and exciton-exciton couplings:
\[
H_{\mathrm{int}} = -\hbar g_i (\sigma_i^\dagger b + \sigma_i b^\dagger) - \hbar g_j (\sigma_j^\dagger b + \sigma_j b^\dagger),
\]
where $g_i$ and $g_j$ denote the coupling strengths between the $i$-th and $j$-th excitons and the plasmon mode ($g_i=g_j=41.7$ meV considered here).
 
The pumping term is included under the assumption that the plasmon mode has a much larger cross-section than the excitons, so only the plasmons are directly driven by the external field:
\[
H_{\mathrm{pump}} = \frac{\mathcal{E}}{2} \left( e^{-i\omega t} b^\dagger + e^{i\omega t} b \right),
\]
where $\hbar \mathcal{E} = \boldsymbol{\mu}_{\mathrm{c}} \cdot \mathbf{E}_0$ is the Rabi frequency associated with the driving laser field $\mathbf{E}_{\mathrm{L}} = \mathbf{E}_0 \cos(\omega t)$, and $\boldsymbol{\mu}_{\mathrm{c}}$ is the effective dipole moment of the nanocavity. In the weak-excitation limit and under the rotating-wave approximation, the total Hamiltonian becomes
\[
H = \Delta_0 \sum_{i,j} \sigma_{i,j}^\dagger \sigma_{i,j} + \Delta_{\mathrm{c}} b^\dagger b  - \sum_{i,j} \hbar g_{i,j} (\sigma_{i,j}^\dagger b + \sigma_{i,j} b^\dagger) + \frac{\mathcal{E}}{2} (b^\dagger + b),
\]
where $\Delta_0 = \omega_0 - \omega$ and $\Delta_{\mathrm{c}} = \omega_{\mathrm{c}} - \omega$ are the detunings of the excitons and the plasmon relative to the laser frequency.

Dissipation is included via decay rates $\Gamma_{\mathrm{c}}=310$ meV and $\Gamma_0=25$ meV \cite{liu_strong_2017} for the plasmon and excitons, respectively. The open-system dynamics are described by the master equation
\[
\dot{\rho} = \mathcal{L}(\rho) = \frac{i}{\hbar} [\rho, H] + \mathcal{L}_{\mathrm{pl}}(\rho) + \mathcal{L}_0(\rho),
\]
where the first term accounts for coherent evolution and the Lindblad superoperators $\mathcal{L}_{\mathrm{c}}$ and $\mathcal{L}_0$ describe dissipation. 
The steady-state solution is obtained numerically using the open-source QuTiP toolbox~\cite{johansson2012qutip}. Once the steady-state density matrix $\rho$ is known, the plasmon population in the cavity $n_c = \langle b^\dagger b \rangle$ and the normalized second-order correlation function
\[
g^{(2)}(\tau) = \frac{\langle b^\dagger(0) b^\dagger(\tau) b(\tau) b(0) \rangle}{\langle b^\dagger(0) b(0) \rangle^2}
\]
can be calculated. 

In the normalized second-order correlation function $g^{(2)}$, the denominator $\langle b^\dagger b \rangle = n_\mathrm{c}$ corresponds to the plasmon population (Fig.~\ref{fig:4}d1), which is proportional to the emission intensity. The spectrum is normalized by the driving field amplitude $\mathcal{E}$ and exhibits a prominent splitting at $\Delta_\mathrm{c} = 0$. From literature, we know that photon-blockade effect relies on the anharmonicity of the Jaynes-Cummings ladder~\cite{manjavacas2012plasmon} is a good way of generating quantum light.  $g^{(2)} < 1$ indicates sub-Poissonian statistics characteristic of nonclassical light. Note that sub-Poissonian statistics are the hallmark of single-photon sources, where the probability of simultaneously emitting two photons is strongly suppressed.  
Specifically, when $g^{(2)}(0) < g^{(2)}(\tau) < 1$, the system exhibits antibunching, meaning the likelihood of emitting two photons simultaneously is lower than for coherent light ($g^{(2)} = 1$). Such a system can function as a single-photon source if $g^{(2)}(0) < 0.5$, with $g^{(2)}(0) = 0$ representing the ideal strong antibunching limit. In contrast, bunched (thermal) photons satisfy $g^{(2)}(0) > g^{(2)}(\tau)$, corresponding to multiphoton emission events. 

As shown in Fig.~\ref{fig:4}d2, both antibunching and bunching photons are observed near the resonance splitting ($\Delta_c = 0$). On resonance, the NPcoM--exciton hybrid system emits antibunched photons with a sub-Poissonian correlation $g^{(2)}(0) = 0.16$, demonstrating well single-photon emission. As the detuning $\Delta_c$ increases, the photon statistics transition to bunching with $g^{(2)}(0) \simeq 1.4$. For large detunings, the emission approaches coherent light again, with $g^{(2)}(0) = 1$. Therefore, the NPcoM--exciton hybrid system provides a fundamental nano-cQED platform for exploring quantum optical phenomena.

In summary, we have demonstrated a NPcoM architecture that deterministically forms sub-nanocavities with deeply subwavelength mode volumes, surpassing conventional NPoM structures. By integrating monolayer TMDC excitons as a probe, the NPcoM platform realizes stronger single-exciton coupling strength, leading to pronounced multibranch Rabi splittings beyond 200 meV and ultrabright polaritonic PL. Remarkably, both the Rabi frequency and its figure of merit ($\Omega/\Gamma \simeq 1$) approach the optimal record for plasmon–monolayer TMDC exciton systems, well above the typical $\Omega/\Gamma \simeq 0.4 \rm\; to\;0.7$ range reported in earlier studies.
In addition, this architecture enables quantum-optical phenomena such as photon blockade at the single-exciton level. Numerical simulations of the two-exciton NPcoM system predict antibunching sub-Poissonian statistics with $g^{(2)}(0) = 0.16$, confirming the potential for single-photon emission. The photon statistics can be tuned from antibunching to bunching by detuning the cavity, demonstrating controllable quantum light generation. Together, these results establish NPcoM sub-nanocavities as a versatile, deterministic nano-cQED platform, offering a unique testbed for exploring quantum optics at the ultimate nanoscale.

\section*{Author Contributions}

H.H. and X.S. contributed equally to this work. 

 \section*{Conflict of interests}
The authors declare no competing interests.

 \section*{Acknowledgments}
This work was supported by the National Natural Science Foundation of China (Grant No. 12204362, 12274334, 62475071 and 52488301) and the National Key Research and Development Program of China (Grant No. 2024YFA1409902).

\bibliography{picocavity1}
\end{document}